 \documentclass[final,5p]{elsarticle}
\usepackage{amssymb}
\journal{arXiv}
\usepackage{amsmath}
\usepackage{amssymb}
\usepackage{psfrag}
\usepackage{color}
\usepackage{stfloats}
\usepackage{fixltx2e}
\usepackage{mathrsfs}
\usepackage{tabularx}
\usepackage{booktabs}
\usepackage{colortbl}
\usepackage{rotating}
\usepackage{boldline}
\usepackage{bm}
\usepackage{changes}
\usepackage{url}
\usepackage{adjustbox}
\usepackage{graphicx}
\usepackage{cellspace}
\usepackage{lineno}

\begin{document}
\renewcommand{\topfraction}{0.98}   
\renewcommand{\bottomfraction}{0.98}
\setcounter{topnumber}{3}
\setcounter{bottomnumber}{3}
\setcounter{totalnumber}{4}         
\setcounter{dbltopnumber}{4}        
\renewcommand{\dbltopfraction}{0.98}
\renewcommand{\textfraction}{0.05}  
\renewcommand{\floatpagefraction}{0.95}      
\renewcommand{\dblfloatpagefraction}{0.95}   
\newcommand{\beq}{\begin{equation}}
\newcommand{\eeq}{\end{equation}}
\newcommand{\divg}{\mbox{\rm{div}}\,}
\newcommand{\Divg}{\mbox{\rm{Div}}\,}
\newcommand{\D}  {\displaystyle}
\newcommand{\DS} {\displaystyle}
\newcommand{\RM}[1]{\textit{\MakeUppercase{\romannumeral #1{}}}}
\newtheorem{remark}{\bf{{Remark}}}
\def\sca   #1{\mbox{\rm{#1}}{}}
\def\mat   #1{\mbox{\bf #1}{}}
\def\vec   #1{\mbox{\boldmath $#1$}{}}
\def\scas  #1{\mbox{{
\footnotesize
{${\rm{#1}}$}}}{}}
\def\scaf  #1{\mbox{{\tiny{${\rm{#1}}$}}}{}}
\def\vecs  #1{\mbox{\boldmath{
\footnotesize
{$#1$}}}{}}
\def\tens  #1{\mbox{\boldmath{
\footnotesize
{$#1$}}}{}}
\def\tenf  #1{\mbox{{\sffamily{\bfseries {#1}}}}}
\def\ten   #1{\mbox{\boldmath $#1$}{}}
\def\Ass  {\overset{\hspace*{0.4cm} n_{\scas{el}}}
          {\underset{\scaf{c},\scaf{d}=1}{\msf{A}}}}
\def\ltr   #1{\mbox{\sffamily{#1}}}
\def\bltr  #1{\mbox{\sffamily{\bfseries{{#1}}}}}
\sloppy
\begin{frontmatter}
\title{\Large 
{\textsf{\textbf{Emergent symmetry in mushroom-based foods}}}}
\author[inst1]{Skyler R. St. Pierre}
\author[inst2]{Thibault Vervenne} 
\author[inst1]{Ethan C. Darwin}
\author[inst1]{Ellen Kuhl\corref{cor1}}
\cortext[cor1]{corresponding author}
\affiliation[inst1]{Department of Mechanical Engineering, Stanford University, 
Stanford, California, United States}
\affiliation[inst2]{Department of Mechanical Engineering,
KU Leuven, Leuven, Belgium}
\begin{abstract} %
Mushroom-based foods exhibit anisotropic fibrous microstructures formed by networks of hyphae and represent a unique class of structured soft matter. These materials provide an opportunity to probe a fundamental mechanics question: does anisotropic structure translate into anisotropic constitutive symmetry? Here we combine directional tension, compression, and shear experiments with automated model discovery and sensory evaluation to investigate three mushroom-based foods with distinct microstructural architectures: mycelium, fruiting body, and a protein-mycelium blend. We show that these three materials exhibit distinct degrees of directional organization, yet span a broad spectrum of constitutive symmetry classes, from strongly anisotropic to effectively isotropic behavior. Sensory evaluation reveals a similar progression, while perceived fibrousness remains largely independent of directional stiffness. Automated model discovery introduces fiber-dependent invariants only when required by the data and directly identifies the governing symmetry class from experiments. Our results show that visual appearance alone does not uniquely determine material symmetry; instead, constitutive symmetry depends on the statistically relevant descriptors of the microstructure. These findings establish mushroom-based foods as a model system to study emergent symmetries in structured soft matter and provide a general framework to discover constitutive symmetry directly from experimental data.\\[4.pt]
Our source code, data, and examples are available at 
https:/\!/github.com/LivingMatterLab/AI4Food.
\end{abstract}
\end{frontmatter}
\noindent{\textbf{\textsf{When structure does not imply symmetry.}}}
Continuum theories often assume that microstructural alignment dictates the symmetry of the mechanical response \cite{Truesdell1965}.
From early theories of rubber elasticity \cite{Treloar1948}
to modern descriptions of fiber-reinforced solids \cite{HolzapfelGO2000}, 
scientists have linked macroscopic behavior to underlying microstructure through invariant-based constitutive laws \cite{Rivlin1948}. 
In this framework, 
\textit{anisotropy} arises when internal architecture introduces preferred directions that break rotational \textit{symmetry} and produce direction-dependent stiffness, strength, and deformation. 
Invariant-based formulations express this behavior through scalar measures \cite{Spencer1971}, 
the total stretch 
$I_1$, 
the shear-like distortion 
$I_2$, and
the volume change
$I_3$
for the isotropic response, and
the fiber stretch squared 
$I_4$ and
the nonlinear fiber stretch 
$I_5$
for the anisotropic response~\cite{Ogden1984}. 
This paradigm motivates the widespread assumption 
that visible structural alignment 
implies anisotropic mechanical symmetry. 
Yet, many soft materials, 
including 
polymers with oriented molecular chains~\cite{Flory1953}, 
cartilage with collagen networks~\cite{Mow1980}, and
brain with axonal fiber bundles~\cite{Franceschini2006}, 
challenge this assumption: 
microstructural organization does not always translate into direction-dependent macroscopic response.\\[-8.pt]

\begin{figure}[b]
\centering
\includegraphics[width=0.48\textwidth]{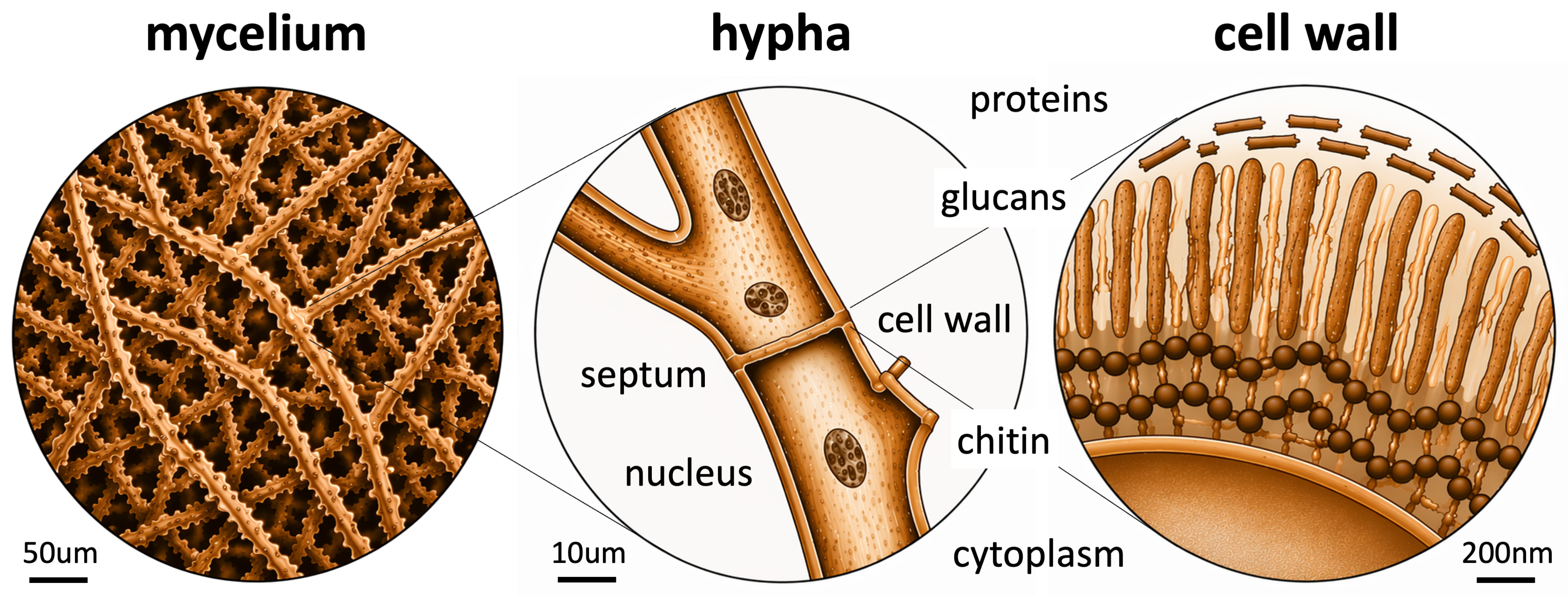}
    \caption{{\noindent\textbf{\textsf{Microstructural hierarchy of fungi-based materials.}}}
Mycelium network composed of interconnected hyphae that form a disordered but locally aligned filamentous architecture (left);
single hypha with septum and intracellular organization, illustrating the tubular geometry and internal lumen (middle);
cell wall ultrastructure with layered organization of chitin, glucans, and proteins (right).
The hierarchical organization across length scales suggests structural anisotropy, yet does not, \textit{a priori}, determine the mechanical symmetry class.
Scale bars: 50\,$\mu$m (mycelium), 10\,$\mu$m (hypha), 200\,nm (cell wall).}
\label{fig01}
\end{figure}
\noindent{\textbf{\textsf{Model system.}}}
Fungal protein materials provide a compelling system to test this assumption~\cite{Assenza2019}. Their microstructure consists of interconnected hyphae that form hierarchical networks with pronounced directional organization across multiple length scales~\cite{Watkinson2016}, from micron-scale filaments to millimeter-scale assemblies (Fig.~\ref{fig01}). These architectures resemble classical fiber-reinforced materials and suggest an inherent capacity for anisotropic mechanical behavior, consistent with established descriptions of fungal ultrastructure and cell wall organization~\cite{Bowman2006}. At the same time, these materials have emerged as promising alternatives to animal meat, where fibrous texture governs sensory perception and consumer acceptance \cite{Finnigan2025}. This dual role places fungal systems at the intersection of soft matter physics, biological materials, and food science. This raises a central question: does anisotropic microstructure necessarily produce anisotropic mechanical response—and does this response control perceived fibrousness? More fundamentally, \textit{does structure dictate symmetry}, or does symmetry only emerge through the response?\\[6.pt]
\noindent{\textbf{\textsf{Experimental strategy.}}}
To address this question, we combine controlled experiments with data-driven model discovery to probe how mechanical symmetry emerges in structured soft materials. We study three fungi-based protein systems derived from distinct structural origins--mushroom root {\it{mycelium}}, the {\it{fruiting body}}, and a {\it{protein mycelium}} blend—which provide a spectrum of microstructural organization and symmetry classes. We prepare samples in two orthogonal orientations relative to the dominant structural direction: in-plane, parallel to the apparent fiber alignment, and cross-plane, perpendicular to it. For each material and direction, we perform $n=10$ tension, compression, and shear experiments up to 10\% strain, which enables direct comparison across loading modes and orientations \cite{StPierre2024}. This results in a total of $n=180$ tests. We extract effective stiffnesses via linear regression in the small-strain regime and compare in-plane and cross-plane responses statistically to quantify anisotropy and test whether structural alignment implies directional invariance.\\[6.pt]
\noindent{\textbf{\textsf{Continuum framework.}}}
We interpret all experiments within a finite-deformation continuum framework that links the microstructural architecture to the macroscopic response through invariant-based constitutive laws. We characterize deformation through the deformation gradient and construct scalar \textit{invariants} that encode the symmetry of the response \cite{Truesdell1965}. 
In isotropic materials, 
the strain energy depends solely on three invariants,
the total stretch $I_1$,  
shear-like distortion $I_2$, and 
volume change $I_3$. 
In anisotropic materials, 
additional invariants
like the fiber stretch squared $I_4$ or 
nonlinear fiber stretch $I_5$ 
introduce preferred directions and reduce symmetry. 
Classical invariant theory 
provides the formal basis for this representation~\cite{Spencer1971}, 
while modern formulations of fiber-reinforced soft materials 
extend it to biological and engineered systems~\cite{Holzapfel2000}. 
In this framework, \textit{the selection of invariants 
determines the effective material symmetry}. 
Here, instead of a priori prescribing the symmetry group, 
we discover symmetry 
through automated model discovery 
using a constitutive neural network \cite{Linka2023}
that expresses the strain energy 
as a sparse combination of invariant-based terms \cite{McCulloch2024}.
\begin{figure}[b]
\centering
\includegraphics[width=0.48\textwidth]{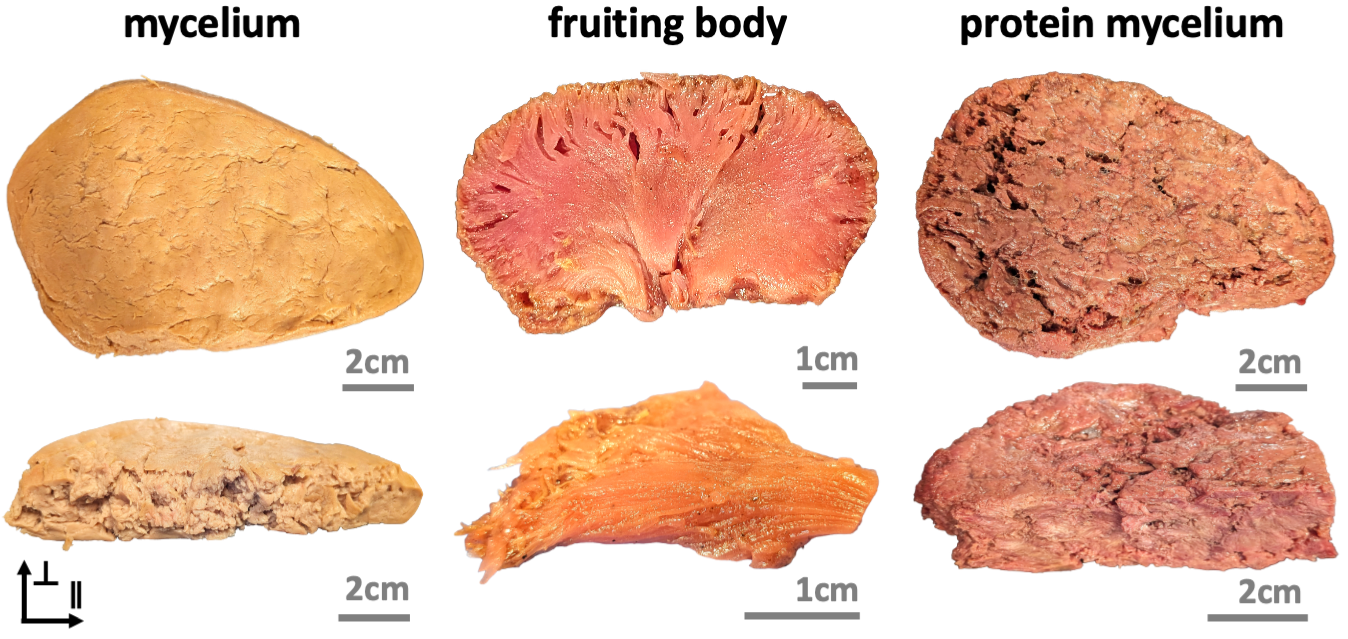}
\caption{{\sffamily{\bfseries{Macroscopic structure of fungi-based materials.}}}
Representative samples of 
mycelium (left), 
fruiting body (middle), and 
protein mycelium blend (right)
in top view (top) and cross-section (bottom). 
All three materials exhibit visually fibrous architectures, 
while the degree of structural alignment and heterogeneity varies across systems.
Scale bars: 1–2\,cm.}
\label{fig02}
\end{figure}
\\[6.pt]
\noindent{\textbf{\textsf{Macrostructure.}}}
At the macroscopic scale, all materials exhibit visually fibrous architectures, primarily in the plane, with distinct degrees of alignment and heterogeneity (Fig.~\ref{fig02}). 
Mycelium appears relatively homogeneous with limited directional features, 
the fruiting body displays pronounced fiber alignment, and 
the protein mycelium blend shows heterogeneous structure with visible local organization. 
The shape of the samples defines the in-plane ($||$) 
and cross-plane ($\perp$) directions.
\begin{figure}[b]
\centering
\includegraphics[width=0.48\textwidth]{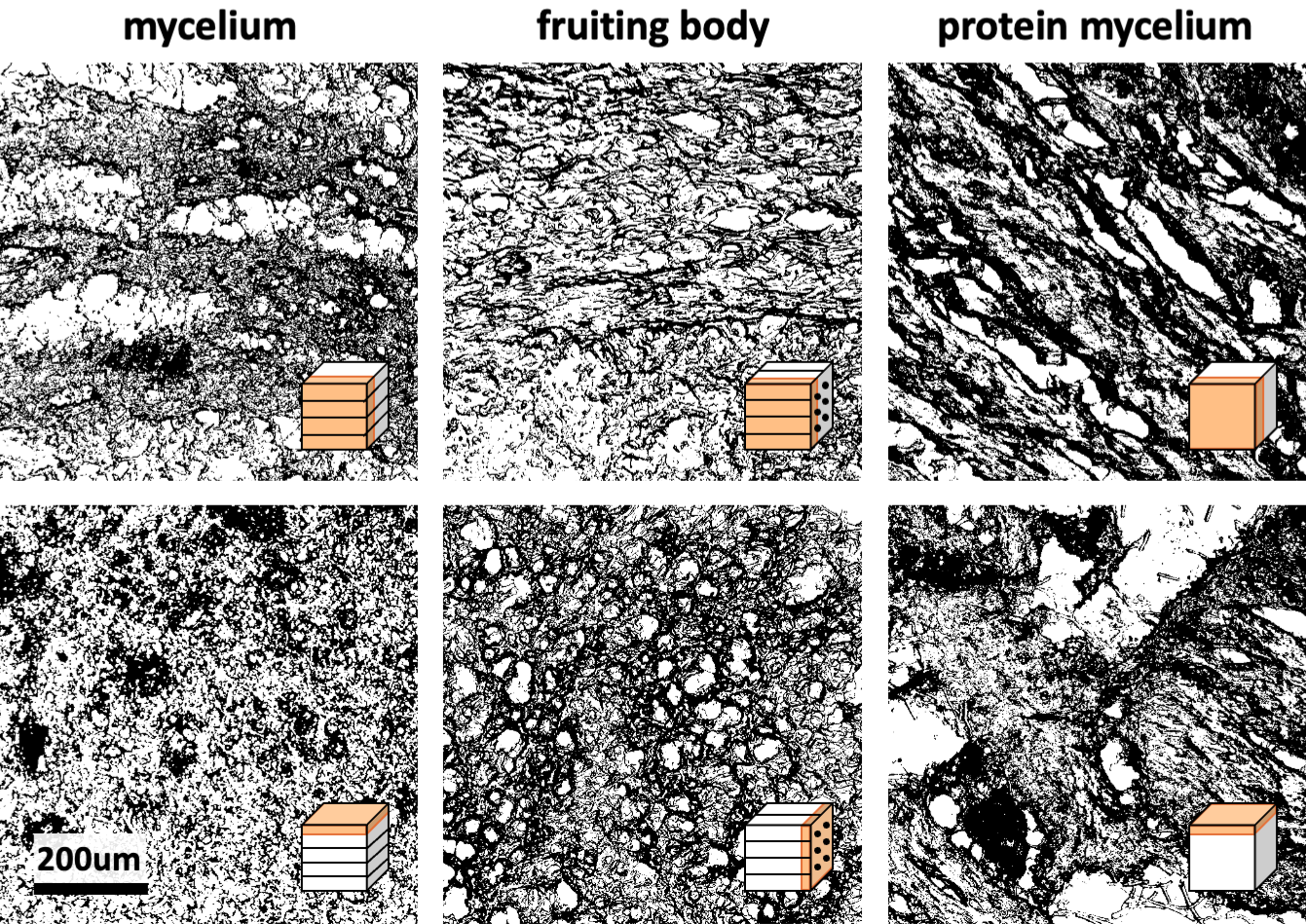}
\caption{{\sffamily{\bfseries{Microscopic structure of fungi-based materials.}}}
Light microscopy images of 
mycelium (left), 
fruiting body (middle), and 
protein mycelium blend (right),
in-plane (top) and cross-plane (bottom). 
Mycelium exhibits large sheet-like structures; 
fruiting body shows densely connected fibers with preferred orientation; and 
protein mycelium blend combines locally aligned regions with large voids. 
Scale bars: 200\,$\mu$m.}
\label{fig03}
\end{figure}
\\[6.pt]
\noindent{\textbf{\textsf{Microstructure.}}}
Light microscopy reveals that the three materials exhibit distinct internal organizations (Fig.~\ref{fig03}). 
Mycelium forms large sheet-like structures with limited alignment, 
the fruiting body consists of densely connected fibers with preferred orientation, and 
the protein mycelium blend combines locally aligned regions with large voids. 
In all cases, structural elements define an apparent fiber direction
that suggests \textit{reduced rotational symmetry} 
and a potential anisotropic response. 
However, the diversity of microstructural motifs raises the question 
to what extent anisotropy at the microscale 
translates into anisotropic macroscopic behavior. \
\begin{figure*}[ht]
\centering
\includegraphics[width=1.0\textwidth]{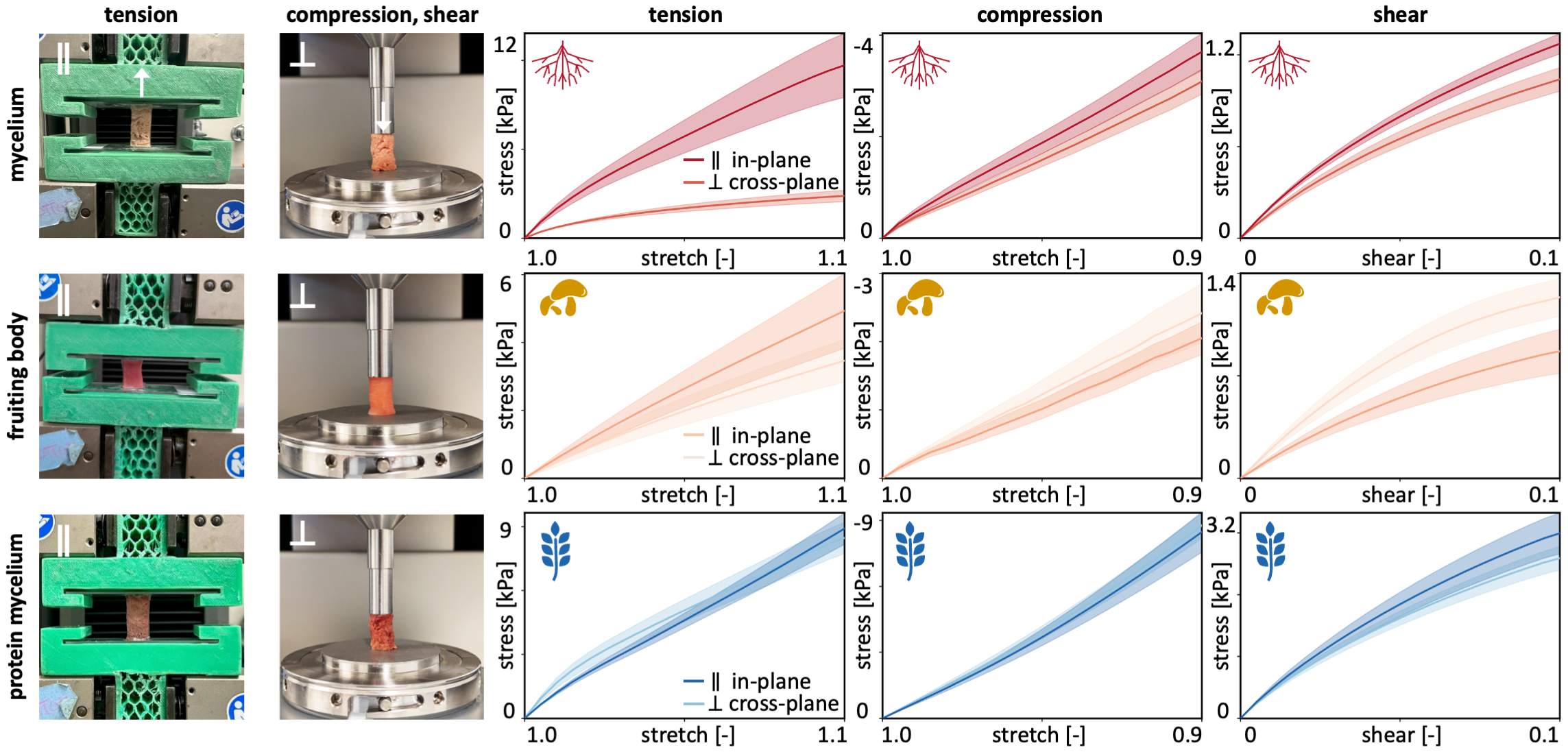}
\caption{{\sffamily{\bfseries{Direction-dependent mechanical response of fungi-based materials.}}}
Tension, compression, and shear tests and
stress–strain curves for 
mycelium (top)
fruiting body (middle), and 
protein mycelium blend (bottom) 
measured in-plane (dark colors, $||$) and cross-plane (light colors, $\perp$) 
(mean $\pm$ s.e.m., $n = 10$). 
Mycelium exhibits strong directional dependence 
with pronounced differences in-plane and cross-plane
across all loading modes; 
fruiting body shows moderate directional differences, most notably in shear; and 
protein mycelium blend displays nearly overlapping responses in both directions, indicating an effectively isotropic mechanical behavior despite its structured appearance.}
\label{fig04}
\end{figure*}
\\[6.pt]
\noindent{\textbf{\textsf{Mechanical response.}}}
Directional mechanical testing directly answers this question and reveals distinct responses (Fig.~\ref{fig04}).
Mycelium exhibits a {\it{strong anisotropy}}, 
with a clear separation between the in-plane and cross-plane behavior 
across all three loading modes. 
Its peak tensile stresses vary from
9.7~kPa in-plane to 2.4~kPa cross-plane—a 
more than fourfold difference. 
The fruiting body shows a {\it{moderate anisotropy}}, 
with smaller directional stress variations,
ranging from
4.9~kPa in-plane to 3.5~kPa cross-plane	in tension
— a difference of less than 50\%.
In contrast, 
the protein mycelium blend
displays a nearly perfect {\it{isotropy}},
with almost identical
in-plane and cross-plane curves across all three loading modes. 
Interestingly, 
the anisotropic mycelium and fruiting body
both display a notable tension-compression asymmetry,
while the isotropic protein mycelium blend
does not.
These results show that similarly complex fibrous microstructures 
can produce markedly different mechanical symmetries.
\begin{figure}[ht]
\centering
\includegraphics[width=\columnwidth]{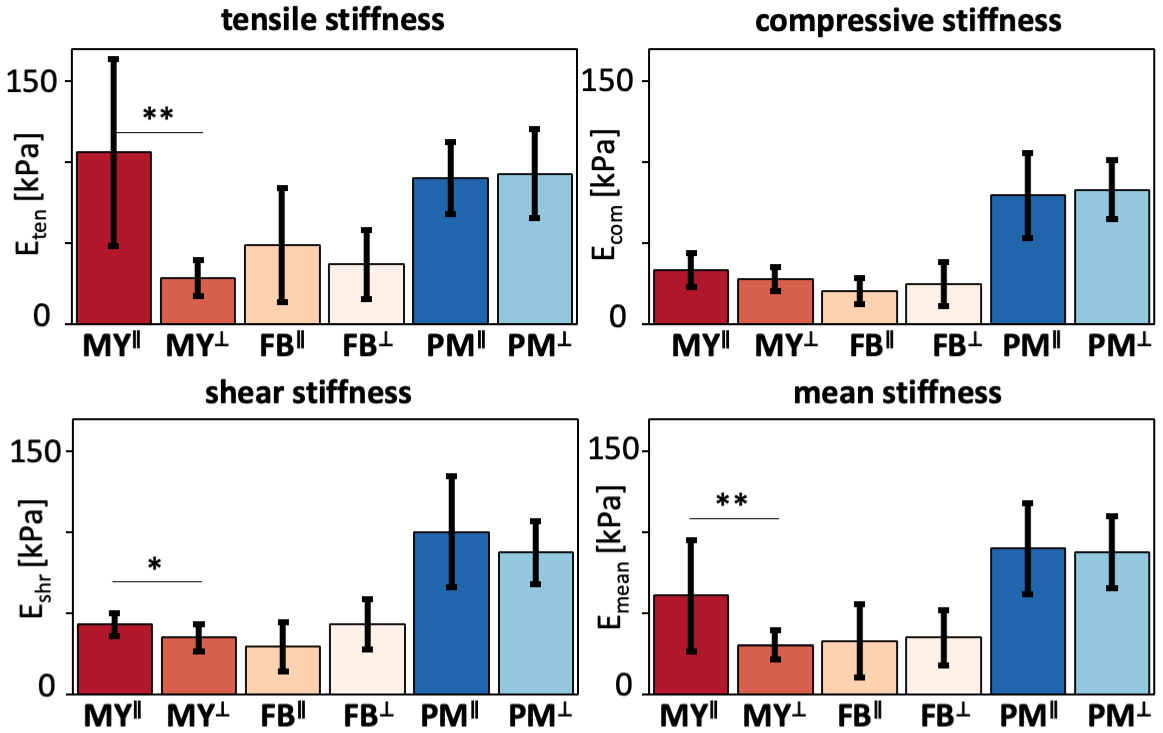}
\caption{{\sffamily{\bfseries{Direction-dependent mechanical stiffness of fungi-based materials.}}}
Mechanical attributes--tensile, compressive, shear, and mean stiffness--for 
mycelium         ({\sffamily{MY}}), 
fruiting body    ({\sffamily{FB}}), and 
protein mycelium ({\sffamily{PM}}) 
measured in-plane ($||$) and cross-plane ($\perp$) 
(mean $\pm$ s.d., * for $p < 0.05$, ** for $p < 0.01$, n = 10). 
Mycelium exhibits significant anisotropy ($p < 0.01–0.05$); 
fruiting body shows anisotropic trends, but not statistically significant; and
protein mycelium blend remains isotropic ($p > 0.05$).}
\label{fig05}
\end{figure}
\\[6.pt]
\noindent{\textbf{\textsf{Mechanical stiffness.}}}
Stiffness measures quantify these differences and establish distinct symmetry classes (Fig.~\ref{fig05}). 
Mycelium exhibits statistically significant 
directional differences ($p<0.01-0.05$),
with the strongest stiffness difference in tension,
106~kPa in-plane versus 29~kPa cross-plane,
which confirms strong anisotropy. 
The fruiting body shows weak or non-significant trends ($p>0.05$), 
which places it in a transitional regime 
between anisotropic and isotropic behavior. 
The protein mycelium blend shows no statistically significant differences 
between in-plane and cross-plane responses ($p>0.05$), 
which confirms isotropic symmetry. 
These results establish a clear progression 
from anisotropic to isotropic symmetry 
across structurally similar materials. 
\begin{figure*}[ht]
\centering
\includegraphics[width=1.0\textwidth]{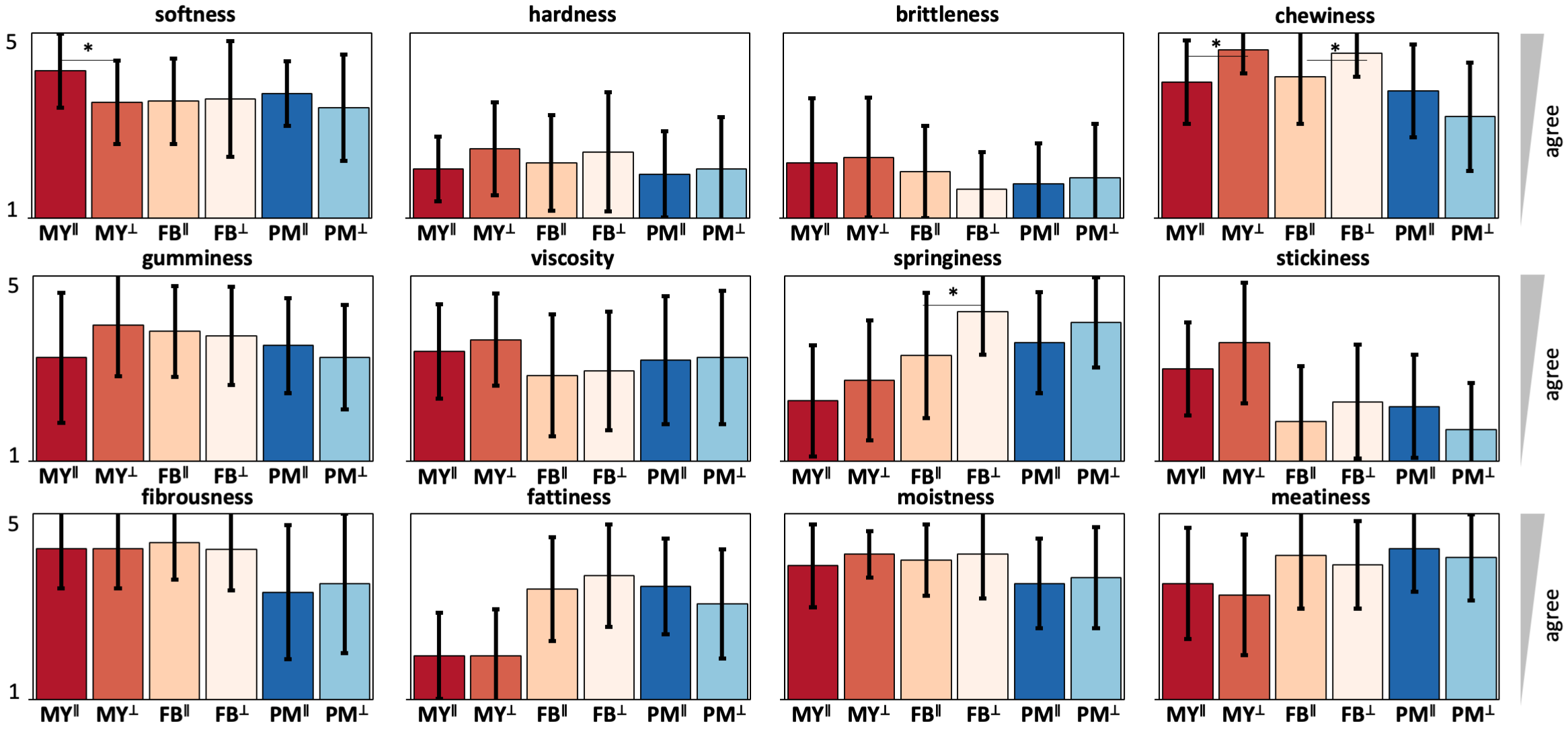}
\caption{{\sffamily{\bfseries{Direction-dependent sensory perception of fungi-based materials.}}}
Sensory attributes--softness, hardness, brittleness, chewiness, gumminess, viscosity, springiness, stickiness, fibrousness, fattiness, moistness, and meatiness--for 
mycelium         ({\sffamily{MY}}), 
fruiting body    ({\sffamily{FB}}), and 
protein mycelium ({\sffamily{PM}}) 
measured in-plane ($||$) and cross-plane ($\perp$) 
(mean $\pm$ s.d., * for $p < 0.05$, *** for $p < 0.001$,
$n = 16$ for {\sffamily{MY}} and {\sffamily{FB}}, $n = 21$ for {\sffamily{PM}}).
Mycelium exhibits anisotropy in softness, hardness, chewiness, and stickiness ($p < 0.05$); 
fruiting body shows anisotropy in chewiness and springiness ($p < 0.05$); and
protein mycelium blend is predominantly isotropic except for chewiness ($p < 0.05$).}
\label{fig06}
\end{figure*}
\\[6.pt]
\noindent{\textbf{\textsf{Sensory perception.}}}
Sensory evaluation provides a complementary perspective (Fig.~\ref{fig06}). Fungal materials are widely regarded as texturally compelling alternatives to animal meat due to their fibrous hyphal microstructure (Fig.~\ref{fig01}).
Mycelium shows attribute-specific anisotropy with largest directional differences in softness, hardness, chewiness, and stickiness ($p < 0.05$). Similarly, the fruiting body shows anisotropy in chewiness and springiness ($p < 0.05$). Protein mycelium blend remains isotropic with no significant differences except in chewiness ($p < 0.05$).
Interestingly, while mycelium exhibits significant directional differences in stiffness, fruiting body and protein mycelium blend do not (Fig.~\ref{fig05}), 
fibrousness shows no statistically significant directional dependence within any product ($p > 0.05$). Perceived fibrousness thus does not directly follow from microstructural alignment or stiffness, but emerges from a combination of mechanical and sensory cues.
This observation agrees with the broader sensory literature, which suggests that the perception of fibrous structures depends on how a food fragments during chewing and subsequently forms a swallowable bolus \cite{Lillford2018}, rather than on directional stiffness alone \cite{Tac2026}. More generally, only limited work has examined the relationship between structural anisotropy and sensory perception in foods \cite{Oppen2024}. This highlights the need for quantitative frameworks that link constitutive mechanics with human perception \cite{Tac2026a}. \\[6.pt]
\noindent{\textbf{\textsf{Model discovery.}}}
Automated model discovery identifies the effective material symmetry directly from the experimental data. 
Mycelium's discovered free energy,
\[
\psi 
= 5.07\,\mbox{kPa} \,[\,I_1 - 3\,] 
+ 1.64\,\mbox{kPa} \, \langle I_5 - 1\rangle^2 \,,
\]
features 
a neo-Hookean isotropic first-invariant term and 
an anisotropic fifth-invariant term,
which confirms the anisotropic nature of mushroom root mycelium. 
The fruiting body's free energy,
\[
\psi 
= 4.42\,\mbox{kPa} \, [\,I_1 - 3\,] 
+ 0.69\,\mbox{kPa} \, [\,\exp\left(0.70\langle I_5 - 1\rangle^2\right) - 1 \,]\,,
\]
consists of 
the same neo-Hookean term, but now paired with
an anisotropic exponential fifth-invariant term,
which, 
at $0.69\,\mbox{kPa}$,
is weighted lower
than the anisotropic term for mycelium,
at $1.64\,\mbox{kPa}$.
The protein mycelium blend's free energy,
\[
\psi =\,14.30~\mbox{kPa} \, [\,I_1 - 3\,]\,,
\]
is entirely neo-Hookean
with \textit{isotropic} symmetry described solely by the first invariant,
which confirms the isotropic nature of the protein mycelium blend.
%
These results show that effective material symmetry 
emerges from mechanical response rather than visual structure. 
The preference for the fifth invariant $I_5$ over the fourth invariant $I_4$
reflects the fact that,
in in-plane tension, $I_5=\lambda^4$ varies more strongly than $I_4=\lambda^2$, and 
in in-plane shear, $I_5=1+\gamma^2$ remains active while $I_4=1$ produces no anisotropic contribution. 
As a result, the discovered model prefers~$I_5$ over~$I_4$. \\[6.pt]
\noindent{\textbf{\textsf{Conclusion.}}}
We show that visual appearance alone does not uniquely determine constitutive symmetry in mushroom-based foods.
Although all three materials exhibit visually similar fibrous architectures, they span a broad spectrum of mechanical symmetry classes, from strongly anisotropic to effectively isotropic behavior. 
Automated model discovery identifies these symmetry classes directly from experimental data and introduces fiber-dependent invariants only when the data warrant their inclusion. 
These findings demonstrate that constitutive symmetry depends on the statistically relevant descriptors of the microstructure rather than on structural appearance alone.
Mushroom-based foods therefore provide a unique platform to study emergent symmetries  in structured soft matter. 
More broadly, this framework enables direct identification of constitutive symmetry from experiments and offers a new route to characterize complex materials when structure alone cannot predict macroscopic behavior.
\section*{{Appendix}}
\noindent
In this study, we test three fungi-based materials: 
{\it{mycelium}}, 
{\it{fruiting body}}, and a 
{\it{protein mycelium}} blend. 
The materials originate from commercially available products (Meati Foods, Boulder, CO; OMNI, New York, NY; Beyond, El Segundo, CA), but we refer to them here solely by their structural origin to emphasize generality.
We test each material
in two orthogonal directions, 
the in-plane direction, parallel to the plane of the steak, and 
the cross-plane direction, perpendicular to the plane (Fig.~2).
\\[6.pt]
\noindent{\textbf{\textsf{Microscopy.}}}
We cut each material in multiple orientations and then freeze the samples in Optimal Cutting Temperature compound (Fisher Scientific, Waltham, MA). We cryosection 40 $\mu$m thin slices and then image the samples with a Leica SP8 microscope (Leica Microsystems, Wetzlar, Germany) with 10$\times$ 
magnification under white light (Fig. \ref{fig03}). 
\begin{figure}[h]
    \centering
    \includegraphics[width=1.0\columnwidth]{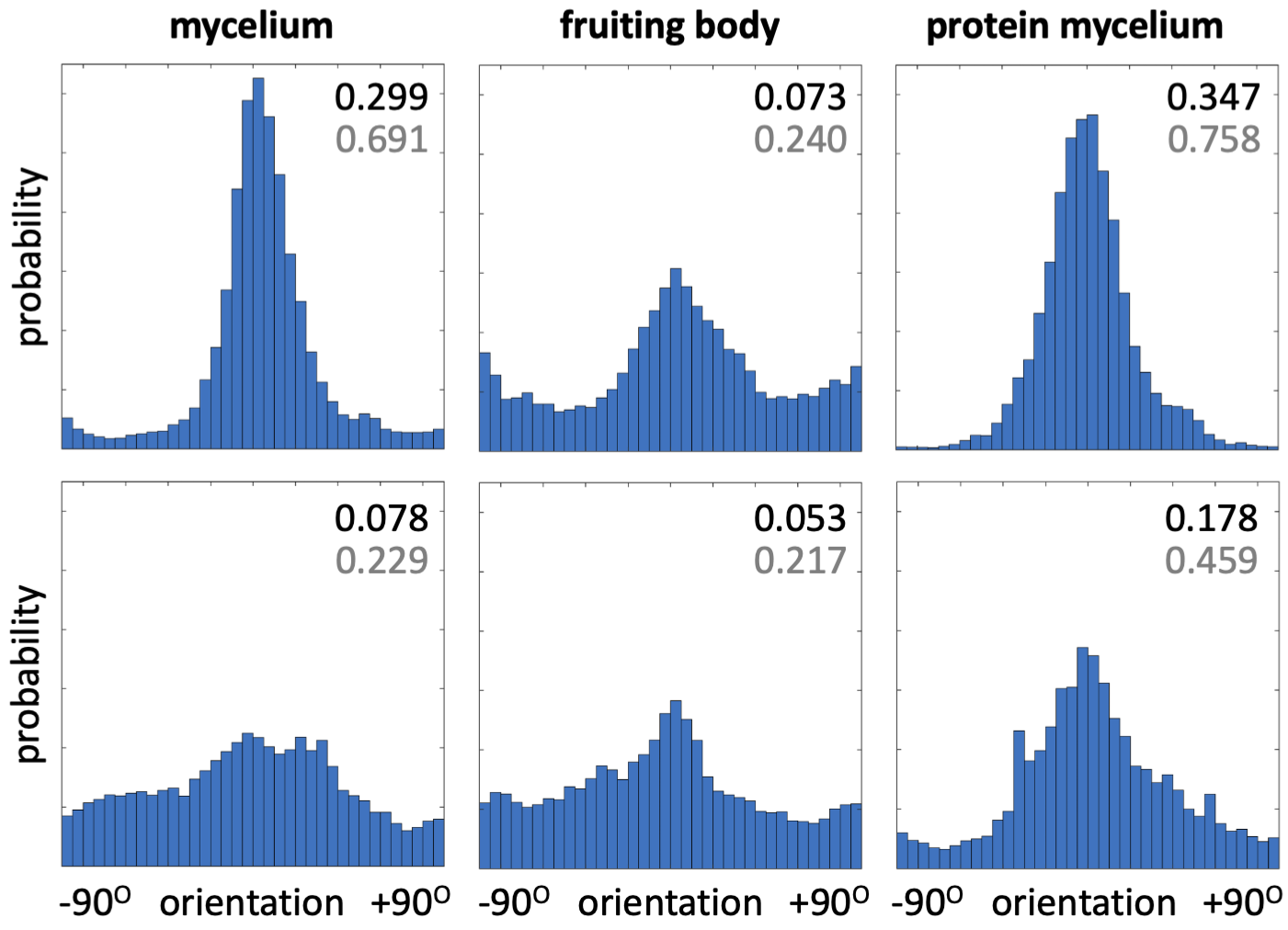}
    \caption{{\sffamily{\bfseries{Orientation distributions of fungi-based materials.}}} Orientation histograms obtained from structure-tensor analysis of the microscopic images (Fig.~\ref{fig03}) for mycelium (left), fruiting body (middle), and protein-mycelium blend (right), in-plane (top) and cross-plane (bottom). Black numbers denote the structure-tensor anisotropy index and quantify the degree of directional organization; gray numbers denote the orientational order parameter and quantify the degree of alignment of the local microstructure. Mycelium and the protein mycelium blend exhibit pronounced in-plane directional organization with high orientational order, while the fruiting body remains nearly isotropic in both planes.}
    \label{fig00sup}
\end{figure}
Quantitative structure-tensor analysis confirms the qualitative observations from the microscopy images. 
Mycelium and protein mycelium blend exhibit the strongest directional organization in plane, with anisotropy indices of 0.299 and 0.347 and orientational order parameters of 0.691 and 0.758 (Fig.~\ref{fig00sup}, left and right, top). 
Protein mycelium blend retains moderate directional organization cross plane, with an anisotropy index of 0.178 and an orientational order parameter of 0.459 (Fig.~\ref{fig00sup}, right, bottom). 
By contrast, the fruiting body exhibits low anisotropy indices, 0.073 in plane and 0.053 cross plane, and low orientational order parameters, 0.240 and 0.213, consistent with a nearly isotropic microstructure (Fig.~\ref{fig00sup}, middle). 
Similarly, mycelium displays reduced directional organization cross plane, with an anisotropy index of 0.078 and an orientational order parameter of 0.229 (Fig.~\ref{fig00sup}, left, bottom).
These results suggest that constitutive symmetry depends on the statistically relevant descriptors of the microstructure, rather than on visual appearance alone.\\[6.pt]
\noindent{\textbf{\textsf{Sample preparation.}}}
For each material and each direction, in-plane and cross-plane, we prepare samples for quasi-static tension, compression, and shear testing with $n = 10$ samples each, resulting in a total of $n = 180$ tests.
For tension testing, we cut rectangular samples of 10$\times$10$\times$20 mm both in-plane, parallel to the apparent fiber alignment, and cross-plane, perpendicular to it.
For compression and shear testing, we use a biopsy punch to extract cylindrical samples of 8\,mm diameter and 10\,mm height, both in-plane and cross-plane.
\\[6.pt]
\noindent{\textbf{\textsf{Tension, compression, and shear testing.}}}
For the tension tests (Fig.~4, first column), we use an Instron 5848 (Instron, Canton, MA) and glue the samples to glass slides, which we mount in custom testing grips \cite{StPierre2024}. 
We mount the samples and extend them to a stretch of $\lambda$ = 1.1  
at a stretch rate of $\dot{\lambda}$ = 0.002/s, 
which translates to a total loading time of $t$ = 50\,s. 
For the compression and shear tests (Fig.~4, second column), we use an HR20 discovery hybrid rheometer (TA Instruments, New Castle, DE) and mount the samples between a 40\,mm diameter base plate and a 8\,mm diameter parallel plate, both sandblasted to avoid slippage \cite{Vervenne2025}. 
For the compression tests, 
we mount the samples and 
compress them to a stretch of $\lambda$ = 0.9  
at a stretch rate of $\dot{\lambda}$ = 0.002/s, 
which translates to a total loading time of $t$ = 50\,s. 
For the shear tests, 
we mount the samples,  
fix them at $\lambda$ = 0.9 compression, and then
shear them to a shear strain of $\gamma$ = 0.1  
at a shear rate of $\dot{\gamma}$ = 0.002/s, 
which translates to a total loading time of $t$ = 50\,s. 
The initial compression is necessary to prevent material slippage.
Across all testing modes, we hold the samples with a small pre-load force between 0.01-0.05 N.
For each material, each direction, and each loading mode,
we calculate the mean and standard error of the mean across all $n = 10$ tests
and illustrate the mean as solid curve and the error as the shaded region around it
(Fig.~4).
The means of the tension, compression, and shear tests serve as the basis for the linear mechanical analysis and for the nonlinear automated model discovery.\\[6.pt]
\noindent{\textbf{\textsf{Mechanical analysis.}}}
To characterize the linear elastic behavior,
we perform a linear regression 
on the quasi-static tension, compression, and shear data 
throughout 10\% tension, 10\% compression and 10\% shear
to extract the compressive, shear, and mean stiffnesses \cite{StPierre2024}.
We postulate a linear stress-strain relation, 
$\sigma = E \cdot \varepsilon$,
and determine the tensile stiffness,
$E_{\rm{ten}}
= (\boldsymbol{\varepsilon} \cdot \boldsymbol{\sigma})
/ (\boldsymbol{\varepsilon} \cdot \boldsymbol{\varepsilon})$,
from the recorded strain-stress pairs
$\{\varepsilon;\sigma\}$
using linear regression.
We use the same method to determine the compressive stiffness
$E_{\rm{com}}$.
Similarly, we postulate a linear shear stress-strain relation, 
$\tau = \mu \cdot \gamma$,
convert the shear modulus $\mu$ into the shear stiffness, 
$E_{\rm{shr}} = 2\, [\,1+\nu\,] \, \mu = 3\, \mu$,
and determine the shear stiffness,
$E_{\rm{shr}} 
= 3\, (\boldsymbol{\gamma} \cdot \hat{\boldsymbol{\tau}})
/ (\boldsymbol{\gamma} \cdot \boldsymbol{\gamma})$,
from the recorded shear strain-stress pairs
$\{\gamma;\tau\}$.
Finally, we average the stiffnesses in tension, compression, and shear
to obtain the mean stiffness, 
$E_{\rm{mean}}= (E_{\rm{ten}}+E_{\rm{com}}+E_{\rm{shr}})/3$.\\[6.pt]
\noindent{\textbf{\textsf{Kinematics.}}}
We analyze all three testing modes combined 
using finite deformation continuum mechanics
\cite{Ogden1984,Truesdell1965}.
During testing, particles 
$\boldsymbol{X}$ of the undeformed sample map to particles, 
$\boldsymbol{x} = \boldsymbol{\varphi}(\boldsymbol{X})$,
of the deformed sample via the deformation map 
$\boldsymbol{\varphi}$.
Similarly, line elements of the 
${\rm{d}}\boldsymbol{X}$ of the undeformed sample map to line elements, 
${\rm{d}}\boldsymbol{x} = \boldsymbol{F} \cdot {\rm{d}} \boldsymbol{X}$, 
of the deformed sample via the deformation gradient, 
$ \boldsymbol{F} 
= \nabla_{\boldsymbol{X}} \boldsymbol{\varphi} 
= \mbox{$\sum_{i=1}^3$}  \;
  \lambda_i \, {\boldsymbol{n}}_{i} \otimes {\boldsymbol{N}}_{i}.$
Its spectral representation introduces the principal stretches $\lambda_i$ and the principal directions 
$\boldsymbol{N}_{i}$ and $\boldsymbol{n}_{i}$
in the undeformed and deformed configurations, 
where $\boldsymbol{F} \cdot \boldsymbol{N}_{i} = \lambda_i \boldsymbol{n}_{i}$.
We adopt a transversely isotropic representation 
as a candidate symmetry class 
and allow the data-driven model discovery 
to determine whether fiber-dependent invariants contribute to the response.
We represent the deformation of the sample through 
three isotropic invariants,
the total stretch
$I_1 = \ten{F} : \ten{F}$,
the shear-like distortion
$I_2 = \frac{1}{2} \; [ I_1^2 - 
      [\, \ten{F}^{\scas{t}} \cdot \ten{F} \,] : 
      [\, \ten{F}^{\scas{t}} \cdot \ten{F} \,] ]$, and
the volume change
$I_3 = \mbox{det} \, (\ten{F}^{\scas{t}} \cdot \ten{F}) = J^2$,
and two fiber direction invariants,
the fiber stretch squared
$I_4 = [\ten{F}^{\scas{t}} \cdot \ten{F}] : \ten{N}$ and
the nonlinear fiber stretch
$I_5 = [\ten{F}^{\scas{t}} \cdot \ten{F}]^2 : \ten{N}$ \cite{Spencer1971,Holzapfel2000}. 
Here, $\ten{N} = \vec{n}_0\otimes \vec{n}_0$ is the structural tensor and 
$\vec{n}_0$ is the unit vector along the microstructural direction (Fig.~\ref{fig01sup}).
We assume that all samples are perfectly incompressible so the 
third invariant always remains equal to one, $I_3 = 1$. 
The remaining four invariants, $I_1$, $I_2$, $I_4$, $I_5$, 
depend on the type of the experiment 
and the direction of testing, in-plane or cross-plane.
The derivatives of the invariants with respect to \ten{F} are
$\partial I_1/\partial \ten{F} = 2\ten{F}$ and
$\partial I_2/\partial \ten{F} = 2[I_1\ten{F}-\ten{F}^{\scas{t}}\cdot\ten{F}]$ 
for the isotropic invariants and
$\partial I_4/\partial \ten{F} = 2\ten{F}\cdot\ten{N}$ and 
$\partial I_5/\partial \ten{F} = 2\ten{F}\cdot[\ten{N}\cdot[\ten{F}^{\scas{t}}\cdot\ten{F}]+[\ten{F}^{\scas{t}}\cdot\ten{F}]\cdot\ten{N}]$ for the anisotropic invariants.
\begin{figure}
    \centering
    \includegraphics[width=0.7\columnwidth]{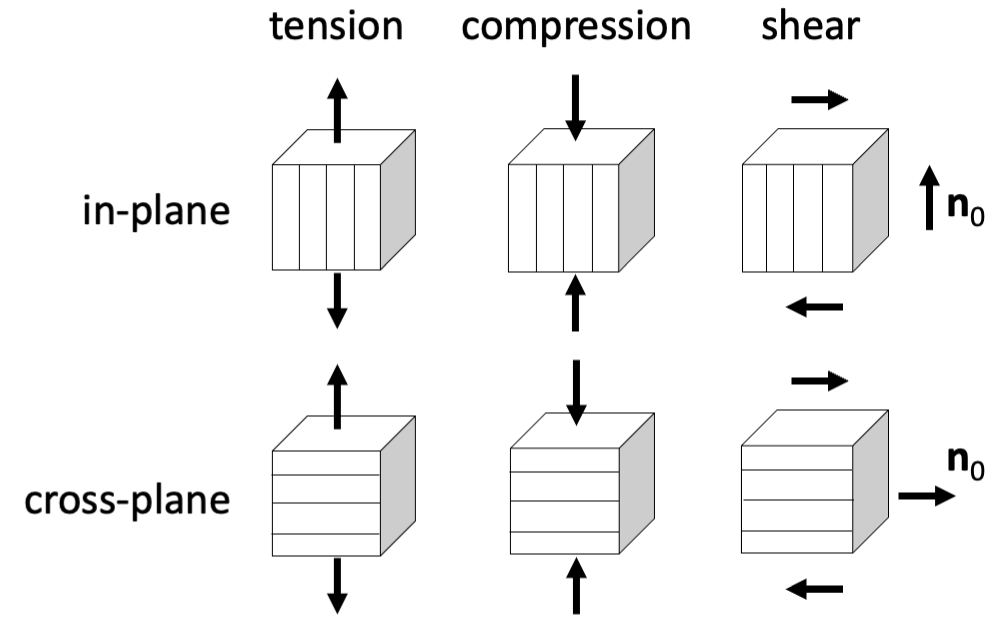}
    \caption{{\sffamily{\bfseries{Microstructure with pronounced fiber direction embedded in isotropic matrix.}}} Tension, compression, and shear loading in the in-plane and cross-plane directions.}
    \label{fig01sup}
\end{figure}
\\[6.pt]
\noindent{\textbf{\textsf{Constitutive equations.}}}
Constitutive equations relate a stress like the Piola or nominal stress $\boldsymbol{P}$, the force per undeformed area that we measure during our experiments, to a deformation measure like the deformation gradient $\boldsymbol{F}$. 
For a hyperelastic material that satisfies 
the second law of thermodynamics,  
we can express the Piola stress,
$ \boldsymbol{P} 
= \partial\psi(\boldsymbol{F})/\partial \boldsymbol{F}
- p \, \boldsymbol{F}^{\text{-t}}$,
as the derivative of the Helmholtz free energy function $\psi(\boldsymbol{F})$ 
with respect to the deformation gradient $\boldsymbol{F}$, 
modified by a pressure term,
$-p \, \boldsymbol{F}^{\text{-t}}$, 
to ensure perfect incompressibility.
Here, the hydrostatic pressure,
$p = - \frac{1}{3} \, \boldsymbol{P}:\boldsymbol{F}$, 
acts as a Lagrange multiplier 
that we determine from the boundary conditions of our experiments. 
Instead of formulating the free energy function directly in terms of the deformation gradient $\psi(\boldsymbol{F})$, we can express it in terms of the invariants, 
$\psi(I_1,I_2, I_4, I_5)$, and obtain the following expression for the stress,
\[
\begin{array}{lcl}
  \boldsymbol{P} 
&=&{\partial\psi}/{\partial I_1} \cdot
  {\partial I_1}/{\partial \boldsymbol{F}} 
+ {\partial\psi}/{\partial I_2} \cdot
  {\partial I_2}/{\partial \boldsymbol{F}}\\ 
&+&{\partial\psi}/{\partial I_4} \cdot
  {\partial I_4}/{\partial \boldsymbol{F}}
+ {\partial\psi}/{\partial I_5} \cdot
  {\partial I_5}/{\partial \boldsymbol{F}}
- p \, \boldsymbol{F}^{\text{-t}}\,.
\end{array}
\]
\noindent{\textbf{\textsf{Automated model discovery.}}}
To characterize the non-linear elastic behavior of each material, we perform automated model discovery to find the transversely isotropic model that best explains the quasi-static behavior for both the in-plane and cross-plane directions. 
\begin{figure}[h]
\centering
\includegraphics[width=\columnwidth]{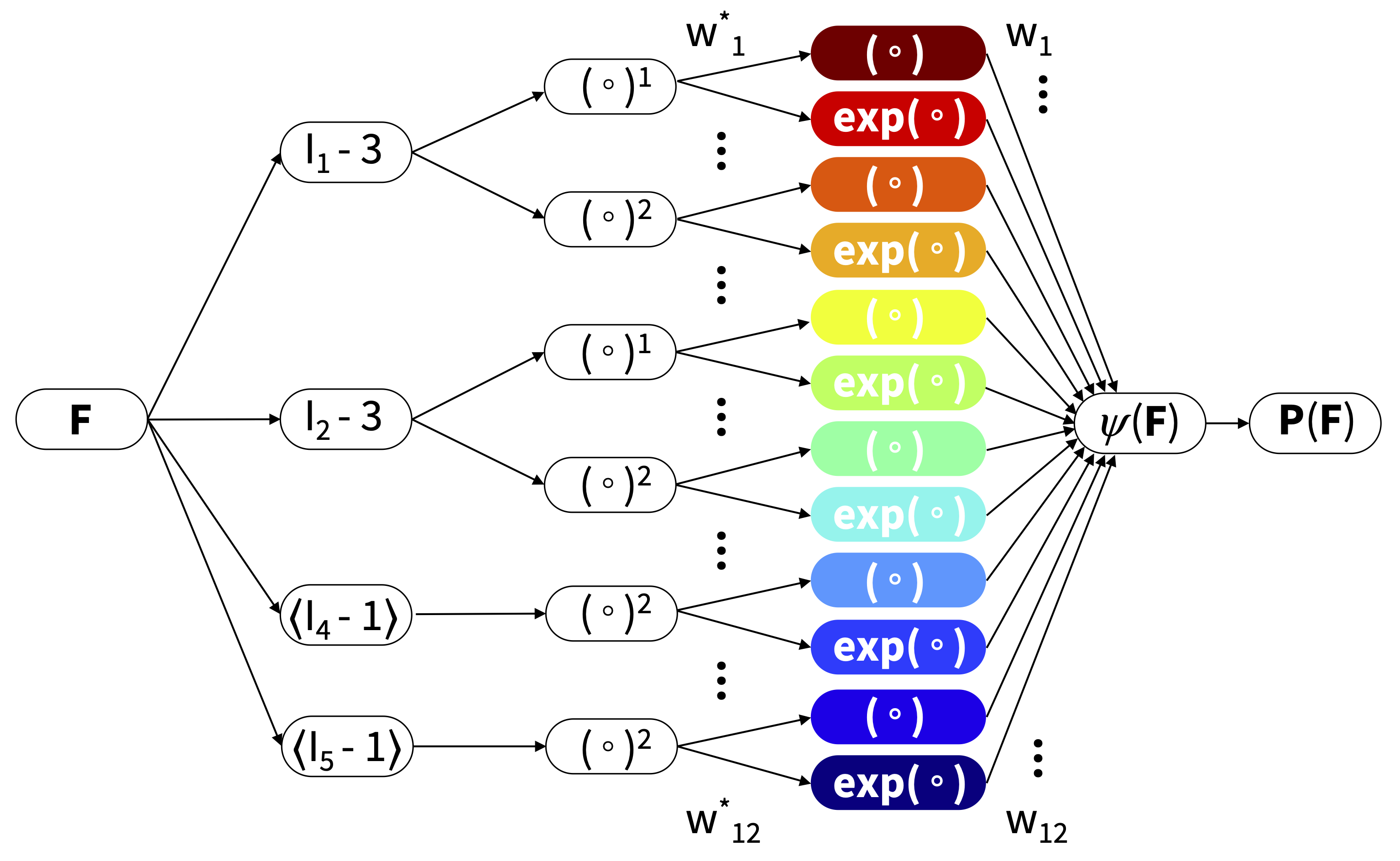}
    \caption{{\sffamily{\bfseries{Automated model discovery.}}}
The constitutive neural network takes the deformation gradient $\ten{F}$ as input and outputs the free energy function $\psi$ from which we calculate the stress 
$\ten{P} = \partial \psi / \partial \ten{F}$.
The network first calculates functions of the invariants,
$I_1$, $I_2$, $I_4$, $I_5$,
and then feeds them into two hidden layers.
The first layer generates the first and second powers, 
$(\circ)^1$ and $(\circ)^2$, of the invariants.
The second layer 
multiplies each term by a weight, $\vec{w^*} = \{ w^*_1, ... w^*_{12} \}$, and
applies the identity and the exponential function, 
$(\circ)$ and $\exp(\circ)$.
The free energy function  $\psi$ is the sum of the resulting 
color-coded terms, multiplied by their weights $\vec{w} = \{ w_1, ... w_{12} \}$. 
We train the network by minimizing the error between model $\ten{P}(\vec{F},\vec{w},\vec{w}^*)$ and data $\hat{\ten{P}}$ 
to learn the network parameters $\vec{w}$ and $\vec{w}^*$,
and apply {\sf{L}}$_1$ regularization to fine-tune the sparsity of the parameter vector $\vec{w}$.}
\label{fig02sup}
\end{figure}
We adopt a twelve-term constitutive neural network (Fig.~\ref{fig02sup})
that takes the first, second, fourth, and fifth invariants, 
$I_1$, $I_2$, $I_4$, $I_5$ of the deformation gradient $\ten{F}$ as input and outputs the free energy function $\psi$ from which we calculate the Piola stress, $\ten{P} = \partial \psi / \partial \ten{F}$ \cite{Linka2023}. We use Macaulay brackets for the fourth and fifth invariants to indicate that these terms only contribute when the fibers are stretched, and not when they are compressed. 
Our network has two hidden layers 
with linear and quadratic functions in the first layer,
$(\,\circ\,)^1$ and $(\,\circ\,)^2$, 
and the identity and exponential functions in the second layer,
$(\,\circ\,)$ and $\exp(\,\circ\,)$ \cite{Linka2023}. 
In total, this network has 
twelve unit-less weights
$\{ w_1^*...w_{12}^* \}$ 
between the first and second layers, and
twelve weights 
$\{ w_1...w_{12} \}$ 
with unit kilopascal out of the second layer. 
As such, it features a total of 
twenty-four network weights that enable $2^{12}$ combinations of terms 
resulting in 4096 possible models. 
We learn the networks weights 
by training the network 
on the tension, compression, and shear data.
The resulting strain-energy functions 
provide compact constitutive descriptions 
over the experimentally investigated deformation range of
10\% tension, compression, and shear.
\\[6.pt]
\noindent{\textbf{\textsf{Tension and compression.}}}
In the tension and compression 
experiments,
we apply a stretch $\lambda =  l / L$, 
that we calculate as the ratio between 
the current and initial sample lengths $l$ and $L$.
We can write the deformation gradient $\boldsymbol{F}$ in matrix representation as
\[
    \boldsymbol{F} =
    \left[ \,\,
    \begin{array}{c@{\hspace*{0.3cm}}cc}
        \lambda & 0 & 0\\
        0 & 1/\sqrt{\lambda} & 0\\
        0 & 0 & 1/\sqrt{\lambda}
    \end{array} 
    \right] 
    \quad \mbox{with} \quad
    \lambda = l/L \, 
    \,.
\]
In tension and compression, the first and second invariants are 
$I_1 = \lambda^2 + 2/{\lambda}$ and $I_2 = 2\lambda  + 1/{\lambda^2}$
and their derivatives along the direction of the applied force are $\partial_{F_{11}} I_1 = 2\lambda$ and $\partial_{F_{11}} I_2 = 4$. Using the zero normal stress condition, $P_{22} = P_{33} = 0$, we obtain the explicit expression for the pressure,
\[ \label{eq:pressure_eq}
   p 
= \frac{2}{\lambda} \frac{\partial \psi}{\partial I_1} 
+ \left[ 2\lambda - \frac{2}{\lambda^2} \right] \frac{\partial \psi}{\partial I_2}\,.
\]
For the {\it{in-plane}} direction, the fourth and fifth invariants are $I_4=\lambda^2$ and $I_5=\lambda^4$ and their derivatives are $\partial_{F_{11}} I_4 = 2\lambda$ and $\partial_{F_{11}} I_5 = 4\lambda^3$. 
Together, this results in the explicit equation for uniaxial stress, 
\begin{equation*}
  P_{11} 
= 2\, \left[ \frac{\partial \psi}{\partial I_1} 
 + \frac{1}{\lambda}\frac{\partial \psi}{\partial I_2} \right]
 \left[ \lambda-\frac{1}{\lambda^2} \right] 
+ 2\lambda \frac{\partial \psi}{\partial I_4} + 4\lambda^3 \frac{\partial \psi} {\partial I_5} \,.
\end{equation*}
For the {\it{cross-plane}} direction, the fourth and fifth invariants are $I_4=1/\lambda$ and $I_5=1/\lambda^2$ and their derivatives are $\partial_{F_{11}} I_4 = 0$. This results in the explicit equation for uniaxial stress, which reduces exactly to the isotropic form with no fiber contributions,
\begin{equation*}
   P_{11} 
 = 2 \left[ \frac{\partial \psi}{\partial I_1} + \frac{1}{\lambda}\frac{\partial \psi}{\partial I_2} \right] \left[\lambda-\frac{1}{\lambda^2} \right] \,.
\end{equation*}
\noindent{\textbf{\textsf{Simple shear.}}}
In the shear experiment, we apply a torsion angle $\phi$,
that translates into the shear stress, $\gamma = r/L \, \phi$,
by multiplying it with the sample radius $r$ and dividing by the initial sample length $L$. We can write the deformation gradient $\ten{F}$ in matrix representation as
\[
    \ten{F} =
    \left[ \,\,
    \begin{array}{c@{\hspace{0.5cm}}c@{\hspace{0.5cm}}c}
        1 & \gamma  & 0 \\
        0 & 1 & 0 \\
        0 & 0 & 1
    \end{array} 
    \right] 
    \quad \mbox{with} \quad
    \gamma = r/L \, \phi
    \,.
\]
In shear, the first and second invariants are 
$I_1 = 3 + \gamma^2$ and $I_2 = 3 + \gamma^2$ and the derivatives in the direction of applied shear are $\partial_{F_{12}} I_1 = 2\gamma$ and $\partial_{F_{12}} I_2 = 2\gamma$.
For the {\it{in-plane}} direction, the fourth and fifth invariants are $I_4=1$ and $I_5=1+\gamma^2$, their derivatives are $\partial_{F_{12}} I_4 = 0$ and $\partial_{F_{12}} I_5 = 2\gamma$, and the explicit equation for simple shear is
\begin{equation*}
  P_{12} 
= 2\gamma \, \left[\frac{\partial \psi}{\partial I_1} + \frac{\partial \psi}{\partial I_2} + \frac{\partial \psi}{\partial I_5} \right] \,. 
\end{equation*}
For the {\it{cross-plane}} direction, the fourth and fifth invariants are $I_4=1+\gamma^2$ and $I_5=(1+\gamma^2)^2+\gamma^2$, their derivatives are $\partial_{F_{12}} I_4 = 2\gamma$ and $\partial_{F_{12}} I_5 = 6\gamma + 4\gamma^3$, and the explicit equation for simple shear is 
\begin{equation*}
    P_{12} 
= 2\gamma \, \left[\frac{\partial \psi}{\partial I_1} + \frac{\partial \psi}{\partial I_2} + \frac{\partial \psi}{\partial I_4}\right] 
+ [\, 6\gamma+4\gamma^3 ]\frac{\partial \psi}{\partial I_5} \,.
\end{equation*}
\noindent{\textbf{\textsf{Loss function.}}}
We minimize the loss function {\sf{L}} that penalizes the error
between the model $\ten{P}(\ten{F}_i,\vec{w},\vec{w}^*)$ and 
the data $\hat{\ten{P}}_i$, 
divided by the number of data points $n_{\rm{data}}$.
We apply {\sf{L}}$_{\rm{1}}$ or lasso regularization \cite{Tibshirani1996},
which supplements the loss function 
by the product of the 
{\sf{L}}$_{\rm{1}}$ norm of the parameter vector $\vec{w}$,
weighted by a penalty parameter $\alpha$,
\[
  {\sf{L}} 
=  \frac{1}{n_{\rm{data}}} \! \sum_{i=1}^{n_{\rm{data}}} \!
|| \ten{P}(\ten{F}_i, \vec{w},\vec{w}^*) - \hat{\ten{P}}_i \, ||^2 
+ \alpha \, || \, \vec{w} \, ||_1 \!
\rightarrow \min_{\vecs{w}}\,.
\]
We select a regularization parameter of $\alpha=0.05$ for all models to prevent overfitting and reduce the number of discovered terms \cite{McCulloch2024,Linka2024}. \\[6.pt]
\noindent{\textbf{\textsf{Sensory survey.}}}
For the sensory survey,
we heat the plain fungi-based steaks,
without spices or condiments,
and prepare bite-sized samples. 
In accordance with our previous studies 
\cite{StPierre2024,Dunne2025,StPierre2025,Vervenne2025},
we recruited $n=16$, $n=21$, and $n=16$ participants 
to participate in the Food Texture Survey to characterize 
{\it{mycelium}} from Mushroom Root Classic Steak (Meati Foods, Boulder, CO), 
{\it{fruiting-body}} from Lion's Mane Mushroom Steak (OMNI, New York, NY), and 
{\it{protein mycelium}} from Beyond Steak Filet (Beyond, El Segundo, CA).
We instruct participants 
to eat samples of the product
and rank its texture features on 
a 5-point Likert scale with twelve questions. 
Each question starts with ``this food is ...'',
following by established sensory texture descriptors
\cite{Nishniari2018,Szczesniak2002}:
{\it{soft}}, {\it{hard}}, {\it{brittle}}, {\it{chewy}}, {\it{gummy}}, {\it{viscous}}, {\it{springy}}, {\it{sticky}}, {\it{fibrous}}, {\it{fatty}}, {\it{moist}}, 
{\it{meaty}}. 
The scale ranges from 5 for strongly agree to 1 for strongly disagree. We provide end-point benchmarks to help participants scale their ratings, e.g., soft = 1 for stale bread and soft = 5 for mashed potatoes, but do not otherwise conduct any training.
The participants 
first eat and rank the steak samples in the cross-plane direction, 
with the fiber direction perpendicular to the chewing direction,
and then in the in-plane direction, 
with the fiber direction parallel to the chewing direction. 
Between the samples, 
we ask participants to cleanse their palate with water to 
minimize residual flavor,
neutralize their taste, and
standardize the sensory environment.
We follow a fixed evaluation protocol consistent with our prior studies~\cite{StPierre2024,Dunne2025,StPierre2025,Vervenne2025} 
to enable direct comparison across datasets.
This research was reviewed and approved 
by the Institutional Review Board at Stanford University 
under the protocol IRB-75418. \\[6.pt]
\noindent{\textbf{\textsf{Statistical analysis.}}}
We analyze our data using custom Python scripts. We use a paired sample t-test for sensory data and Welch's t-test for stiffness data to compare the means in-plane and cross-plane. \\[6.pt]
\noindent{\textbf{\textsf{Mycelium.}}}
Figure \ref{fig03sup} summarizes the results of the automated model discovery to characterize the non-linear elastic behavior of the mycelium material in the in-plane and cross-plane directions for tension, compression, and shear up to 10\%.
\begin{figure*}[ht]
\centering
\includegraphics[width=0.95\textwidth]{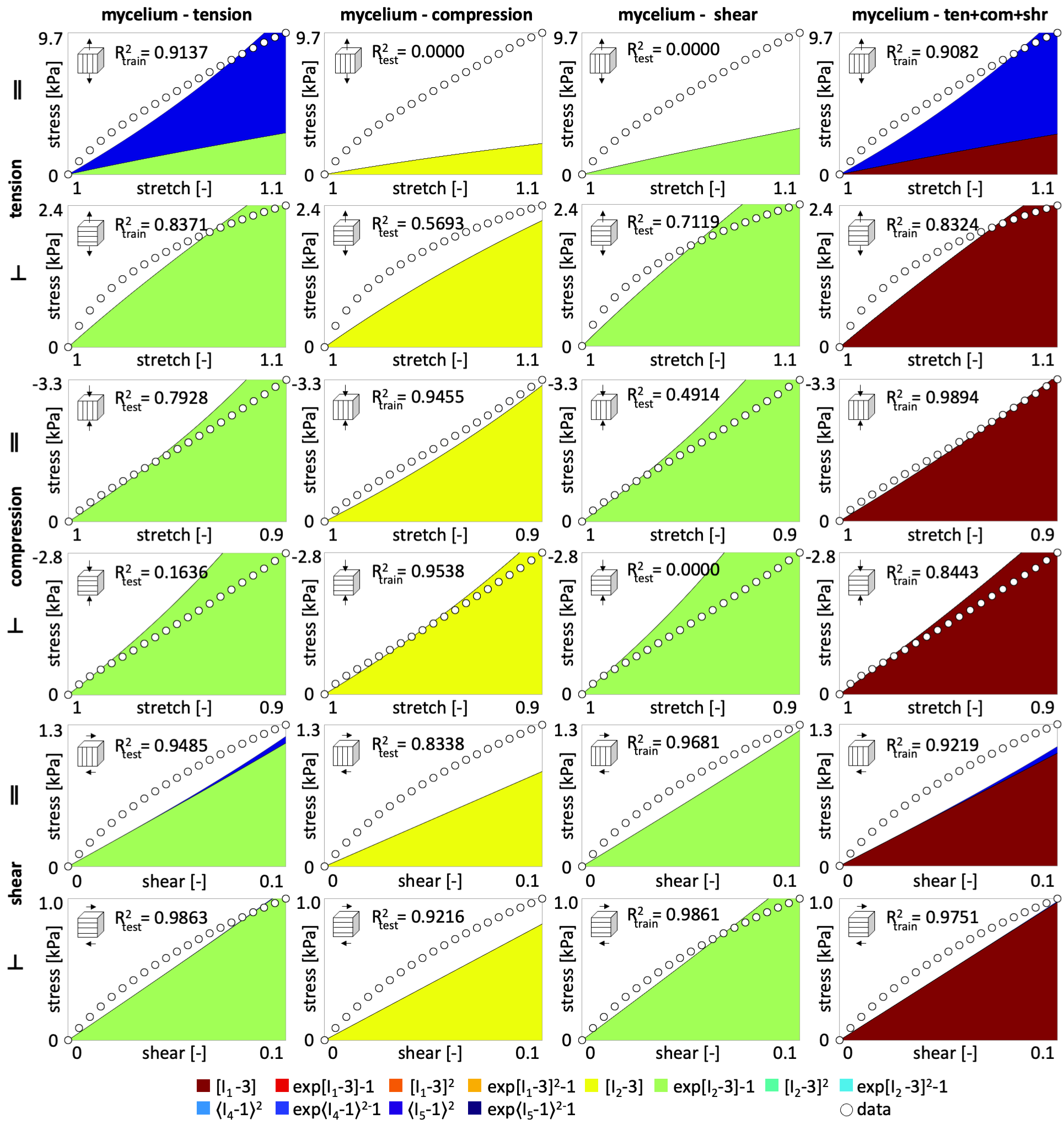} 
\caption{{\sffamily{\bfseries{Discovered models for mycelium.}}} We train the transversely isotropic constitutive neural network in Fig. \ref{fig02sup} with tension, compression, shear, and all three loading modes simultaneously and apply $L_1$-regularization. Pictograms show the fiber orientation, in-plane ($\|$) or cross-plane ($\perp$), and loading mode. The dots indicate the experimental data and the color-coded regions designate the contributions of the twelve model terms to the stress function according to Fig. \ref{fig02sup}. The coefficient of determination, $R^2$, reports the goodness of fit. Mean $R^2 = 0.9119$ for combined model training.}
\label{fig03sup}
\end{figure*}
The dots indicate the experimental data of the quasi-static tension, compression, and shear tests from the average of $n=10$ samples each, the colored regions highlight the contributions of the individual model terms, and the $R^2$ values quantify the goodness of fit.
The transversely isotropic constitutive neural network discovers a two-term model for training on tension only that consists of an isotropic term and a fiber term, 
$\psi = w_6 \, [\,\exp (\, w_6^* \, [I_2-3]) -1] + w_{11} w_{11}^* \langle I_5-1 \rangle ^2$. In contrast, for training on compression only the discovered model is simply $\psi =  w_5 w_5^* [I_2-3]$ and for shear only the discovered model is $ \psi = w_6 \, [\,\exp (\, w_6^* \, [I_2-3]) -1]$. Training with tension only or compression only discovers models that are moderately generalizable to the testing data with average $R^2_{\rm{mean}} = 0.7737$ and $0.7040$. The model trained on shear only struggles to fit tension and compression with an $R^2_{\rm{mean}} = 0.5263$.
Notably, simultaneous training on tension, compression, and shear discovers the linear neo Hookean term and the fifth invariant squared term. This combined model also has the highest goodness of fit across all six in-plane and cross-plane experiments, with $R^2_{\rm{mean}} = 0.9119$. The inclusion of the fifth invariant indicates that mycelium has a preferred fiber direction. 
The discovered parameters for the combined tension, compression, and shear model are $w_1^*=2.2503$, $w_1=2.2540$\,kPa, $w_{11}^*=1.2790$, and $w_{11}=1.2800$\,kPa. 
Taken together, the discovered strain energy function for the mycelium is
\begin{equation*}
    \psi = 5.07 \,\rm{kPa}\, [\,I_1-3\,]+1.64 \,\rm{kPa}\, \langle I_5-1 \rangle ^2 \,.
\end{equation*}
From the high average $R^2$ values, we conclude that this two-term model with the isotropic linear neo Hookean term and the anisotropic nonlinear fiber term provides an excellent fit to the data.  \\[6.pt]
\noindent{\textbf{\textsf{Fruiting body.}}}
Figure \ref{fig04sup} summarizes the results of the automated model discovery to characterize the non-linear elastic behavior of the fruiting body material in the in-plane and cross-plane directions for tension, compression, and shear up to 10\%.
\begin{figure*}[ht]
\centering
\includegraphics[width=0.95\textwidth]{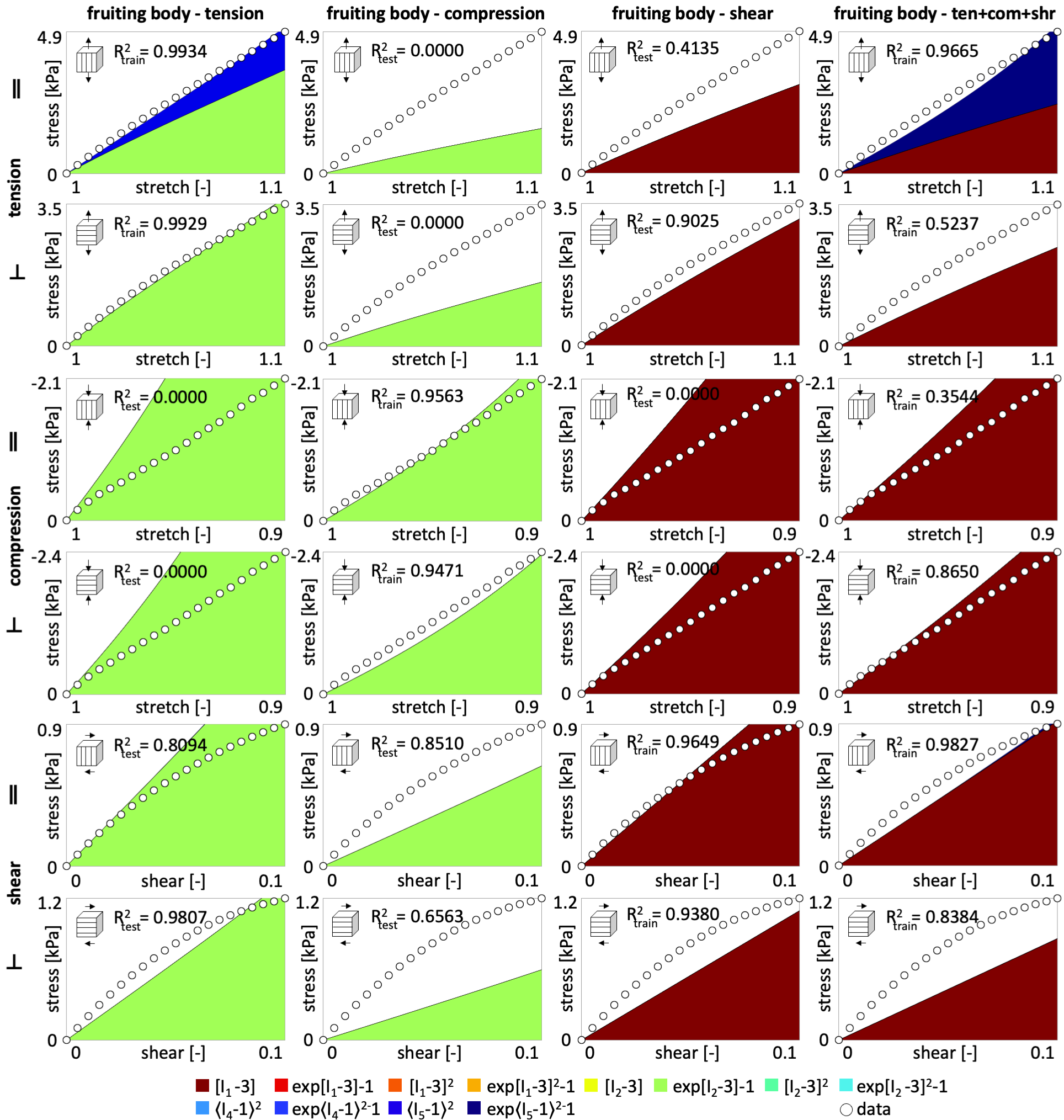} 
\caption{{\sffamily{\bfseries{Discovered models for fruiting body.}}} We train the transversely isotropic constitutive neural network in Fig. \ref{fig02sup} with tension, compression, shear, and all three loading modes simultaneously and apply $L_1$-regularization. Pictograms show the fiber orientation, in-plane ($\|$) or cross-plane ($\perp$), and loading mode. The dots indicate the experimental data and the color-coded regions designate the contributions of the twelve model terms to the stress function according to Fig. \ref{fig02sup}. The coefficient of determination, $R^2$, reports the goodness of fit. Mean $R^2 = 0.7551$ for combined model training.}
\label{fig04sup}
\end{figure*}
The dots indicate the experimental data of the quasi-static tension, compression, and shear tests from the average of $n=10$ samples each, the colored regions highlight the contributions of the individual model terms, and the $R^2$ values quantify the goodness of fit.
Interestingly, the transversely isotropic constitutive neural network discovers nearly the same model when training tension only versus compression only, 
$ \psi = w_6 \, [\,\exp (\, w_6^* \, [I_2-3]) -1]$. In tension training, the model additionally discovers the fiber term, $[w_{11}w_{11}^*\langle I_5-1 \rangle^2]$. However, tension training is unable to predict compression, and compression training is unable to predict tension, with $R^2=0$ for both in-plane and cross-plane testing data.
In contrast, training shear only results in a similar model, the linear neo Hooke model, $ \psi = w_1 w_1^* \, [I_1-3]$, as training on tension, compression, and shear simultaneously. The simultaneous model has an additional the fiber term $w_{12} \, [\exp (w_{12}^* \, \langle I_5-1 \rangle ^2)-1]$, that predominantly contributes to the fit of the tension in-plane data. The goodness of fit across all six tension, compression, and shear in-plane and cross-plane data ranges from $R^2=0.3544$ for compression in-plane to $R^2=0.9827$ for shear in-plane with an average fit of $R^2_{\rm{mean}}=0.7551$
The discovered parameters for the combined tension, compression, and shear model are $w_1^*=2.1712$, $w_1=2.0377$\,kPa, $w_{12}^*=0.7002$, and $w_{12}=0.6881$\,kPa. 
Taken together, the discovered strain energy function for the fruiting body is
\begin{equation*}
    \psi = 4.42 \,\rm{kPa}\, [I_1-3] + 0.69\,\rm{kPa}\, [\exp (0.70 \langle I_5-1 \rangle ^2)-1] \,.
\end{equation*}
\begin{figure*}[ht]
\centering
\includegraphics[width=0.95\textwidth]{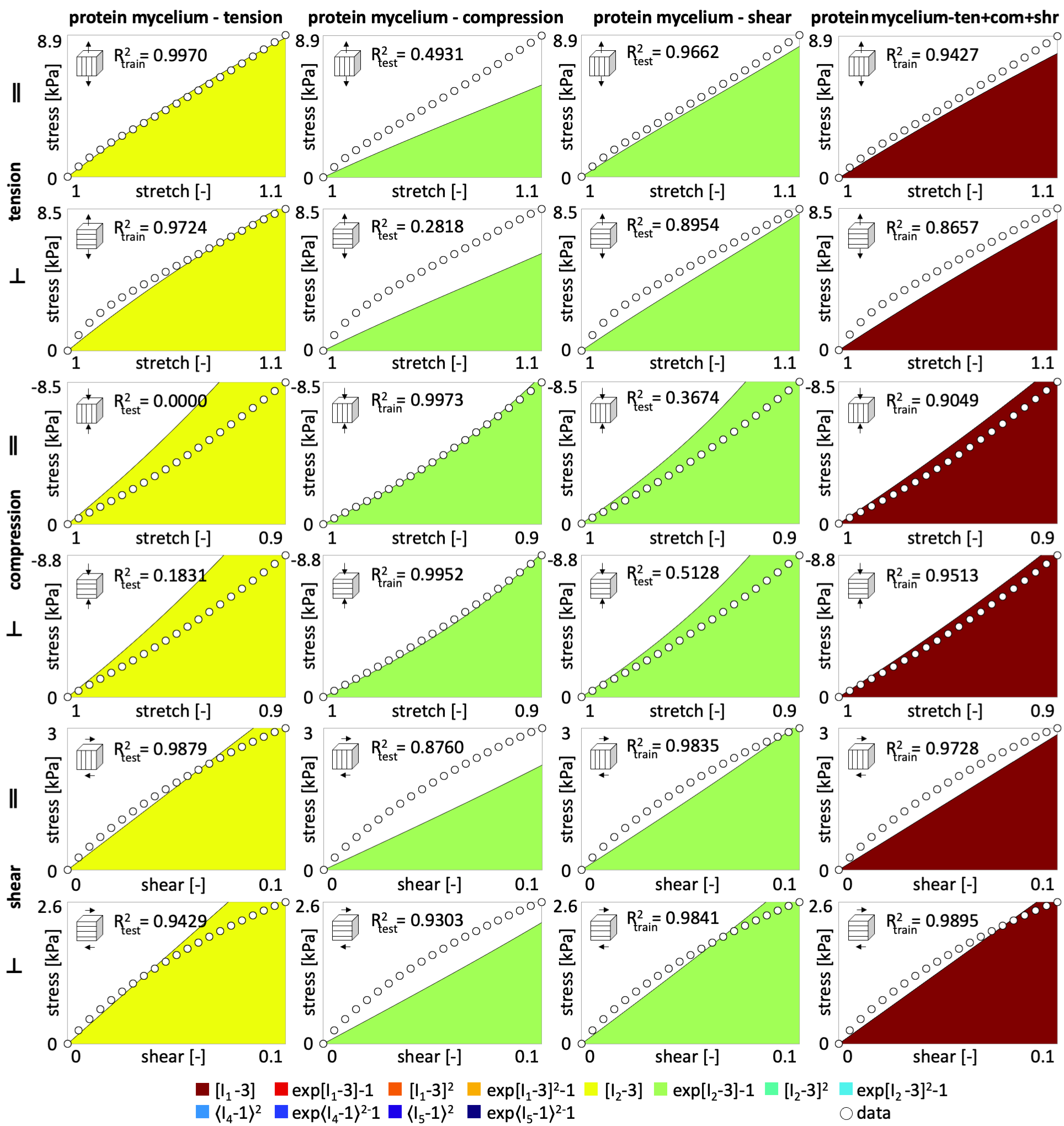} 
\caption{{\sffamily{\bfseries{Discovered models for protein mycelium blend.}}} We train the transversely isotropic constitutive neural network in Fig. \ref{fig02sup} with tension, compression, shear, and all three loading modes simultaneously and apply $L_1$-regularization. Pictograms show the fiber orientation, in-plane ($\|$) or cross-plane ($\perp$), and loading mode. The dots indicate the experimental data and the color-coded regions designate the contributions of the twelve model terms to the stress function according to Fig. \ref{fig02sup}. The coefficient of determination, $R^2$, reports the goodness of fit. Mean $R^2 = 0.9378$ for combined model training.}
\label{fig05sup}
\end{figure*}
From the high average $R^2$ values, we conclude that this two-term model with the isotropic linear neo Hookean term and the anisotropic exponential fiber term provides a reasonable fit to the data. \\[6.pt]
\noindent{\textbf{\textsf{Protein mycelium blend.}}}
Figure \ref{fig05sup} summarizes the results of the automated model discovery to characterize the non-linear elastic behavior of the  protein mycelium blend material in the in-plane and cross-plane directions for quasi-static tension, compression, and shear up to 10\%.
The dots indicate the experimental data of the quasi-static tension, compression, and shear tests from the average of $n=10$ samples each, the colored regions highlight the contributions of the individual model terms, and the $R^2$ values quantify the goodness of fit.
Notably, the transversely isotropic constitutive neural network discovers a one-term model for training on each loading mode. The discovered tension only training model is $\psi = w_5 w_5^* \, [I_2-3]$. For training on compression only and shear only, the discovered model is simply $\psi =  w_6 \, [\,\exp (\, w_6^* \, [I_2-3]) -1]$. All three individual training modes discover models that generalize moderately to the testing data with average $R^2_{\rm{mean}} = 0.6806, 0.7623, 0.7849$ for tension only, compression-only, and shear only.
Interestingly, simultaneous training on tension, compression, and shear discovers the linear neo Hookean model, $\psi = w_1 w_1^* [I_1-3]$, in contrast to the non-linear models discovered by the other three training modes. This combined model also has the highest goodness of fit across all six in-plane and cross-plane experiments, with $R^2_{\rm{mean}} = 0.9378$. The fourth and fifth invariants are not discovered for any training mode for the protein mycelium blend material. 
The discovered parameters for the combined tension, compression, and shear model are $w_1^*=4.1216$ and $w_1=3.4689$\,kPa. 
Taken together, the discovered strain energy function for the protein mycelium blend is
\begin{equation*}
    \psi = 14.30 \,\rm{kPa}\,[\,I_1-3\,] \,.
\end{equation*}
From the high average $R^2$ values, we conclude that this one-term model in terms of only the classical linear neo Hookean term provides an excellent fit to the data.  
\section*{Data availability}
\noindent
Our source code, data, and examples are available at
https://github.com/LivingMatterLab/AI4Food.

\section*{Acknowledgments}
\noindent
The authors thank the Chaudhuri lab 
for use of their cryostat and microscope.
This research was supported by the 
NSF Graduate Research Fellowship,
Stanford DARE Fellowship, 
Research Foundation Flanders FWO Fellowship SB1SE2123N, 
Stanford Bio-X Graduate Student Fellowship,
Stanford Doerr School of Sustainability Accelerator, 
Stanford Bio-X Snack Grant, 
NSF CMMI grant 2320933, and 
ERC Advanced Grant 101141626. 

\section*{Declaration of competing interest}
\noindent
The authors declare that they have no known competing financial interests or personal relationships that could have appeared to influence the work reported in this paper.



\begin{thebibliography}{99}

\bibitem{Truesdell1965}
C. Truesdell and W. Noll,
\textit{The Non-Linear Field Theories of Mechanics}
(Springer, 1965).

\bibitem{Treloar1948}
L.~R.~G. Treloar,
Trans. Faraday Soc. \textbf{44}, 592 (1948).

\bibitem{HolzapfelGO2000}
G.A. Holzapfel, T.C. Gasser, and R.W. Ogden, 
J. Elast. \textbf{61} 1–48 (2000).

\bibitem{Rivlin1948}
R.S. Rivlin,
Phil. Trans. Royal Soc. London A \textbf{240} 459–490 (1948).

\bibitem{Spencer1971}
A.~J.~M. Spencer,
\textit{Theory of Invariants},
(in: Continuum Physics, Academic Press, 1971).

\bibitem{Ogden1984}
R.~W. Ogden,
\textit{Non-Linear Elastic Deformations}
(Dover, 1984).

\bibitem{Flory1953}
P.J. Flory,  
\textit{Principles of Polymer Chemistry}
(Cornell University Press, 1953).

\bibitem{Mow1980}
V.C. Mow, M.H. Holmes, and W.M. Lai,
J. Biomech. \textbf{17}, 377 (1984).

\bibitem{Franceschini2006}
G. Franceschini, D. Bigoni, P. Regitnig, and G.A. Holzapfel,
J. Biomech. \textbf{39}, 1448 (2006).

\bibitem{Assenza2019}
S. Assenza, R. Mezzenga,
Nature Rev. Phys. \textbf{1}, 551-566 (2019).

\bibitem{Watkinson2016}
S.C. Watkinson, L. Boddy, and N.P. Money,
\textit{The Fungi}, 3rd ed.
(Academic Press, 2016)

\bibitem{Bowman2006}
S.M. Bowman and S.J. Free, 
BioEssays. \textbf{28} 799–808 (2006).

\bibitem{Finnigan2025}
T.J.A. Finnigan, H.E. Theobald, and B. Bajka, 
Annu. Rev. Food Sci. Technol. \textbf{16}, 105–25 (2025).

\bibitem{StPierre2024}
S.R. St. Pierre \textit{et al.},
npj Sci. Food. \textbf{8}, 94 (2024).

\bibitem{Holzapfel2000}
G.A. Holzapfel,
\textit{Nonlinear Solid Mechanics}
(Wiley, 2000).

\bibitem{Linka2023}
K. Linka and E. Kuhl, 
Comp. Meth. Appl. Mech. Eng. \textbf{403} 115731 (2023).

\bibitem{McCulloch2024}
J.~A. McCulloch \textit{et al.},
Int. J. Num. Meth. Eng. \textbf{125} e7481 (2024).

\bibitem{Lillford2018}
P. Lillford, 
J. Text. Stud. \textbf{49} 213-218 (2018).

\bibitem{Tac2026}
V. Tac \textit{et al.},
Foods. \textbf{15} 2026 (2026).

\bibitem{Oppen2024}
D. Oppen \textit{et al.},
Crit. Rev. Food Sci. Nutr. \textbf{64} 1158-1176 (2024).

\bibitem{Tac2026a}
V. Tac \textit{et al.}, 
npj Sci. Food. \textbf{10} 199 (2026).

%
%
%

\bibitem{Linka2024}
K. Linka and E. Kuhl, 
Extreme Mech. Lett. \textbf{70} 102181 (2024).

\bibitem{Dunne2025}
R.A. Dunne \textit{et al.},
Food Res Int. \textbf{205} 115876 (2025).

\bibitem{Nishniari2018}
K. Nishinari and Y. Fang, 
J. Text. Stud. \textbf{49} 160-201 (2018).

\bibitem{StPierre2025}
S.R. St. Pierre \textit{et al.},
Curr. Res. Food Sci \textbf{10} 101080 (2025).

\bibitem{Szczesniak2002}
A.S. Szczesniak, 
Food Qual. Pref. \textbf{13} 215-225 (2002).

\bibitem{Tibshirani1996}
R. Tibshirani,
J. R. Statist. Soc. B. \textbf{58} 267-288 (1996).

\bibitem{Vervenne2025}
T. Vervenne, S.R. St Pierre, N. Famaey and E. Kuhl,
Acta Biomat. \textbf{202} 341-351 (2025). 

\end{thebibliography}
\end{document}